\documentclass[12pt]{aastex63}

\usepackage{rotating}
\usepackage{longtable}
\usepackage{graphicx}
\usepackage{txfonts}
\usepackage{verbatim}
\usepackage{enumitem} 

\usepackage{appendix}
\usepackage{hyperref}
\usepackage{nameref}
\usepackage{natbib,twoopt}
\usepackage{color}
\usepackage{threeparttablex}

\defcitealias{Desomma2021}{DS21}
\defcitealias{Desomma2020apj}{DS20b}
\defcitealias{Desomma2020mnras}{DS20a}
\defcitealias{Desomma2022}{DS22}

\received{day month 2025}
\revised{2025}
\accepted{2025}
\submitjournal{ApJ}
\shorttitle{CC Age Distribution in M31 and M33}
\shortauthors{ }
\graphicspath{{./}{figures/}}
\begin{document}

\title{Spatial Age Distribution of Classical Cepheids in Spiral Galaxies: The Cases of M31 and M33}
\correspondingauthor{Giulia De Somma}


\author{Giulia De Somma}
\affiliation{ INAF-Osservatorio astronomico di Capodimonte \\
Via Moiariello 16 \\
80131 Napoli, Italy}
\affiliation{INAF-Osservatorio Astronomico d'Abruzzo \\
Via Maggini sn \\
64100 Teramo, Italy\\}
\affiliation{Istituto Nazionale di Fisica Nucleare (INFN) - Sez. di Napoli\\
Compl. Univ.di Monte S. Angelo, Edificio G, Via Cinthia \\
I-80126, Napoli, Italy}

\author{Marcella Marconi}
\affiliation{ INAF-Osservatorio astronomico di Capodimonte \\
Via Moiariello 16 \\
80131 Napoli, Italy}

\author{Vincenzo Ripepi}
\affiliation{ INAF-Osservatorio astronomico di Capodimonte \\
Via Moiariello 16 \\
80131 Napoli, Italy}

\author{Santi Cassisi}
\affiliation{INAF-Osservatorio Astronomico d'Abruzzo \\
Via Maggini sn \\
64100 Teramo, Italy\\}
\affiliation{Istituto Nazionale di Fisica Nucleare (INFN) - Sezione di Pisa\\
Universit\'a di Pisa, Largo Pontecorvo 3\\
56127 Pisa, Italy\\}

\author{Roberto Molinaro}
\affiliation{ INAF-Osservatorio astronomico di Capodimonte \\
Via Moiariello 16 \\
80131 Napoli, Italy}

\author{Ilaria Musella}
\affiliation{ INAF-Osservatorio astronomico di Capodimonte \\
Via Moiariello 16 \\
80131 Napoli, Italy}

\author{Teresa Sicignano}
\affiliation {European Southern Observatory, \\ Karl-Schwarzschild-Strasse 2, 85748 Garching bei München, Germany\\}
\affiliation{Scuola Superiore Meridionale, \\ Largo S. Marcellino 10\\
80138 Napoli, Italy \\}
\affiliation{ INAF-Osservatorio astronomico di Capodimonte \\
Via Moiariello 16 \\
80131 Napoli, Italy}
\affiliation{Istituto Nazionale di Fisica Nucleare (INFN) - Sez. di Napoli\\
Compl. Univ.di Monte S. Angelo, Edificio G, Via Cinthia \\
I-80126, Napoli, Italy}

\author{Erasmo Trentin}
\affiliation{ INAF-Osservatorio astronomico di Capodimonte \\
Via Moiariello 16 \\
80131 Napoli, Italy}

\begin{abstract}
\noindent


Classical Cepheids can be used as age indicators due to well-established period-age and period-age-color relations. \citet[][]{Desomma2021} refined these relations by including a metallicity term and different Mass-Luminosity assumptions.

In this study, we apply the period-age-metallicity relation for the first time to samples of Classical Cepheids in M31 and M33. For both galaxies, we consider cepheid coordinates and spatial distribution, along with the metallicity gradient by \citet[][]{Zaritsky1994} and \citet[][]{Magrini2007} to provide a metallicity estimate for each pulsator.
Therefore, by applying the period-age-metallicity relation, we derive the individual ages of each Cepheid.

By combining the individual ages and spatial distributions of Classical Cepheids in M31 and M33, we built detailed age maps for both galaxies. Our analysis confirms a radial age gradient, with younger Cepheids preferentially found toward the galactic center. In M31, we confirm an outer ring at $\sim$11 kpc, consistent with previous studies, and identify for the first time an inner ring at $\sim$7 kpc, possibly associated with star formation episodes.

Comparing age gradients at different angles, we find a consistent general trend of ages increasing monotonically with radius. At the same time, we observe smaller-scale differences, particularly in the 90°–180° quadrant, suggesting asymmetric star formation and possible dynamical influences. In contrast, M33 displays a steeper global age gradient, indicating a higher concentration of young stars toward its center.

This study highlights the utility of Cepheids as stellar population tracers, providing insights into the star formation and dynamical evolution of spiral galaxies. Future works will extend this methodology to additional galaxies.

\end{abstract}


\section{Introduction} \label{sec:intro}
Classical Cepheids (or Type I Cepheids) are widely recognized for their pivotal role as primary distance indicators due to the correlation between their pulsation periods and intrinsic luminosities, known as the Period-Luminosity relation (PL). This relation, along with its refinements that include color and metallicity corrections, forms the basis of extragalactic distance measurements \citep[see e.g.][and reference therein]{Bono2024, Breuval2022, Riess2021, Riess2022, Ripepi2020, Ripepi2023}. Unlike Type II Cepheids (BL Herculis, W Virginis and RV Tauri stars), which are older, lower-mass stars, Classical Cepheids (CCs) are young, massive stars (with initial masses between approximately 3 and 13 solar masses) that primarily reside in spiral arms and star-forming regions. Their effective temperatures range from about 5000 K to 7000 K, and their ages span from approximately 20 to 200 million years.

Beyond their role in cosmic distance measurements, CCs also follow a well-established relation between their pulsation periods and stellar ages, known as the Period-Age relation \citep[see e.g.][]{Anderson2016, Bono2005, Efremov1978, EfremovElmegreen1998, Efremov2003, Magnier1997, Senchyna2015, Desomma2020mnras, Desomma2021}. This relation is derived from the combination of pulsation models and stellar evolutionary tracks in the Hertzsprung-Russell diagram, allowing CCs to serve as tracers of stellar population ages and chemical evolution. The theoretical framework for these relations has been progressively refined, with the latest models \citep[][hereinafter DS21]{Desomma2021} providing metal-dependent Period-Age and Period-Age-Color relations, also accounting for variations in the Mass-Luminosity (ML) relation \citep[see e.g.][]{BonoTornambe2000, CasSal2011, SalCas2005}. This refined approach enables a more detailed analysis of CCs across different astrophysical environments.

In this study, we apply the updated period-age-metallicity (PAZ) relation from \citetalias{Desomma2021}, derived by combining pulsation models from \citet[hereinafter DS20b, DS22]{Desomma2020apj, Desomma2022} with the latest BaSTI-IAC evolutionary tracks\footnote{The BaSTI-IAC stellar model repository can be found at: http://basti-iac.oa-abruzzo.inaf.it} \citep[][]{Hidalgo2018}, to two prominent Local Group spiral galaxies: M31 (Andromeda) and M33 (Triangulum). These galaxies, with large CC populations and well-investigated metallicity gradients (\citet [][for M31]{Zaritsky1994} and \citet[][for M33]{Magrini2007}), offer a robust framework for testing PAZ relation and mapping spatial age distributions. M31, the most massive member of the Local Group, provides a unique opportunity to study the age distribution of observed CCs in a spiral galaxy very similar to the Milky Way by relying on datasets such as the PAndromeda survey \citep{Kodric2013, Kodric2018}, which provides the largest CC catalog for the galaxy. For M33, we use the CC sample from the M33 Synoptic Stellar Survey \citep{Pellerin2011}, which offers comprehensive photometric coverage of the galaxy’s disk and a refined period determination.

Following up on the application of the \citet[hereinafter DS20a]{Desomma2020mnras} period-age relation and the \citetalias{Desomma2021} PAZ relation to a sample of Galactic Cepheids using Gaia Data Release 2 (DR2; \citealt{Ripepi2019}) and Early Data Release 3 (EDR3; \citealt{Ripepi2020}), this work presents the first application of the PAZ relation to the M31 and M33 galaxies. By combining this relation with CC spatial distributions, we derive detailed age maps and explore their implications for star formation histories and structural components, such as disks and spiral arms. 

The structure of the paper is as follows: Section 2 details the adopted samples of CCs in M31 and M33.  Section 3 describes the application of PAZ relation and the derivation of the individual ages, for two assumptions on the CC ML relation. Section 4 presents the age map distributions for CCs in M31 and M33. Finally, the discussion summarizes the main results and discusses potential future developments.

\section{Data and Age Estimation Method}

\subsection {The Adopted Classical Cepheid Samples in M31 and M33} 

The M31 CC sample used in this study comes from the PAndromeda project, a Pan-STARRS1 (PS1) survey, covering the full extent of M31 with nightly cadence observations over a $\sim$ 2-3 year baseline in the r-band, later transformed to I-band, and a high temporal resolution for precise period determination, essential for applying the PAZ relation. \citet{Kodric2018} provided the largest catalog, listing 1662 fundamental-mode (F-mode) and 307 first-overtone mode (FO-mode) Cepheids.\footnote{Classical Cepheids pulsate either in the F or FO mode. F-mode CCs exhibit longer periods and larger amplitudes, while FO-mode CCs have shorter periods and smaller amplitudes. This distinction is important because FO-mode Cepheids are systematically brighter than F-mode CCs at the same period, requiring separate PL calibrations. Accurate mode identification is therefore essential to ensure the reliability of age determinations using the PAZ relation.\\
In addition to F-mode and FO-mode pulsators, a small fraction of CCs can also pulsate in the second-overtone mode (SO-mode). However, the number of models in our dataset is insufficient for deriving a reliable Period-Age (PA) relation, and therefore, they are not included in our analysis.}

Regarding the completeness of the sample, we note that the lowest period in the dataset is about two days, which suggests that shorter-period (fainter) CCs are likely missing. \citet[][]{Kodric2013} did not perform a completeness analysis; however, they used the differential imaging technique to detect CCs. Therefore, we expect a good completeness level ($\gtrsim$ 80--90\%) down to P $\sim$ 2 days, even in the most central regions of the galaxy.

For M33, the CC sample originates from the M33 Synoptic Stellar Survey \citep{Pellerin2011}, which combined B, V, and I-band photometry from the DIRECT project with additional follow-up observations from the WIYN 3.5m telescope. The dataset spans most of M33’s disk, with regular observations over a 2-year baseline, albeit with a lower temporal resolution than M31. Despite this, it ensures wide spatial coverage while minimizing crowding biases.

Since the authors did not distinguish CCs in M33 by pulsation mode, we performed this classification using the PL relation in the I band, defining F-mode CCs as those that satisfy the following criterion which has been applied directly to the PL diagram of M33 CCs\footnote{To define the separation between the two pulsation modes, we plotted the density distributions of F-mode and FO-mode CCs in the PL plane and traced the line at the position where the density reaches a minimum between these two populations.}:

\[
I_{\text{mag}} > -3.53 \cdot (\log_{10}(\text{Period}) - 0.30) + 21.91
\]

\noindent
CCs that do not obey this criterion were classified as FO-mode CCs. Following this approach, our final M33 CC sample includes 506 F-mode CCs and 57 FO-mode CCs.

Concerning completeness, as in the case of M31, the lowest period in the M33 CC sample is about 2 days. However, for this galaxy, \citet[][]{Pellerin2011} carried out completeness experiments, finding a 90 \% completeness limit at I = 21.5 mag (corresponding to P $\sim$ 2 days) with no significant spatial variation. This means that we can consider the M33 CC sample essentially complete down to P $\sim$ 2 days.

Finally, we note that for both M31 and M33, the lower limit of P $\sim$ 2 days implies that, in terms of age, we are possibly missing the oldest and likely more external CCs in these galaxies, which we will examine in the upcoming sections.

\subsection{The Individual Age Derivation} 

In this section, we applied the PAZ relation, in the form $\log t$ = $a$ + $b \log P$ + c [Fe/H], from \citetalias[][]{Desomma2021}, adopting a canonical (case A) or a non-canonical (case B) ML relation  \footnote{By canonical luminosity (case A), we refer to models computed without core convective overshooting, rotation, or mass loss, following the ML relation in Section 4 of \citet{BonoTornambe2000}. On the other hand, the non-canonical ML (case B), as discussed in \citetalias[][]{Desomma2020apj, Desomma2022}, corresponds to the case A ML but with the luminosity increased by $\Delta\log(L/L_\odot) = 0.2$ dex. This assumption allows us to mimic the effect of moderate core convective overshooting (or mass loss or rotation). For further details on this topic, we refer to \citetalias[][]{Desomma2020apj, Desomma2022}.}, to the selected M31 and M33 CC samples described earlier, estimating the individual ages for all CCs in these datasets.

\begin{figure*}[ht]
    \hbox{
    \includegraphics[width=0.5\linewidth]{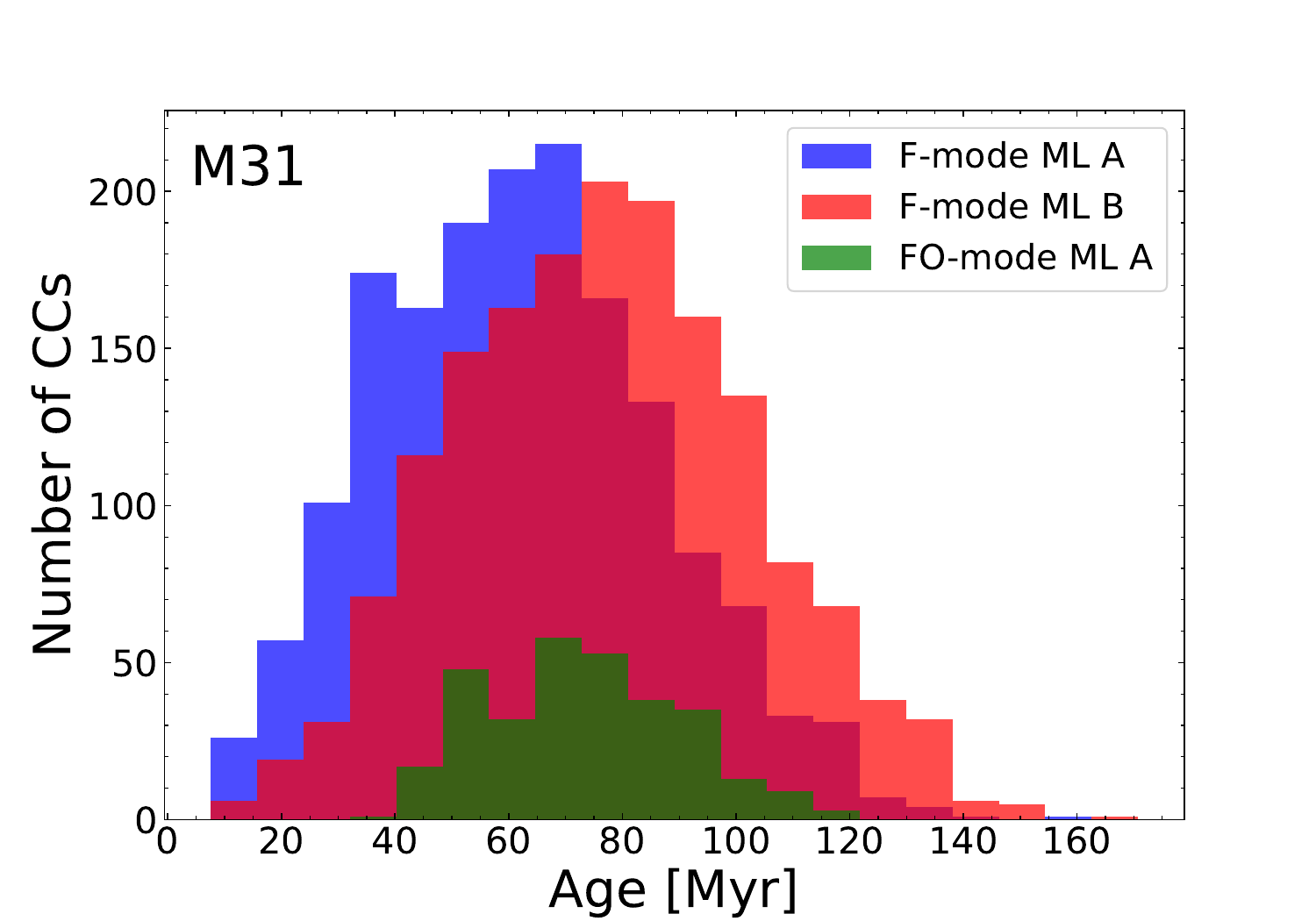}
    \includegraphics[width=0.5\linewidth]{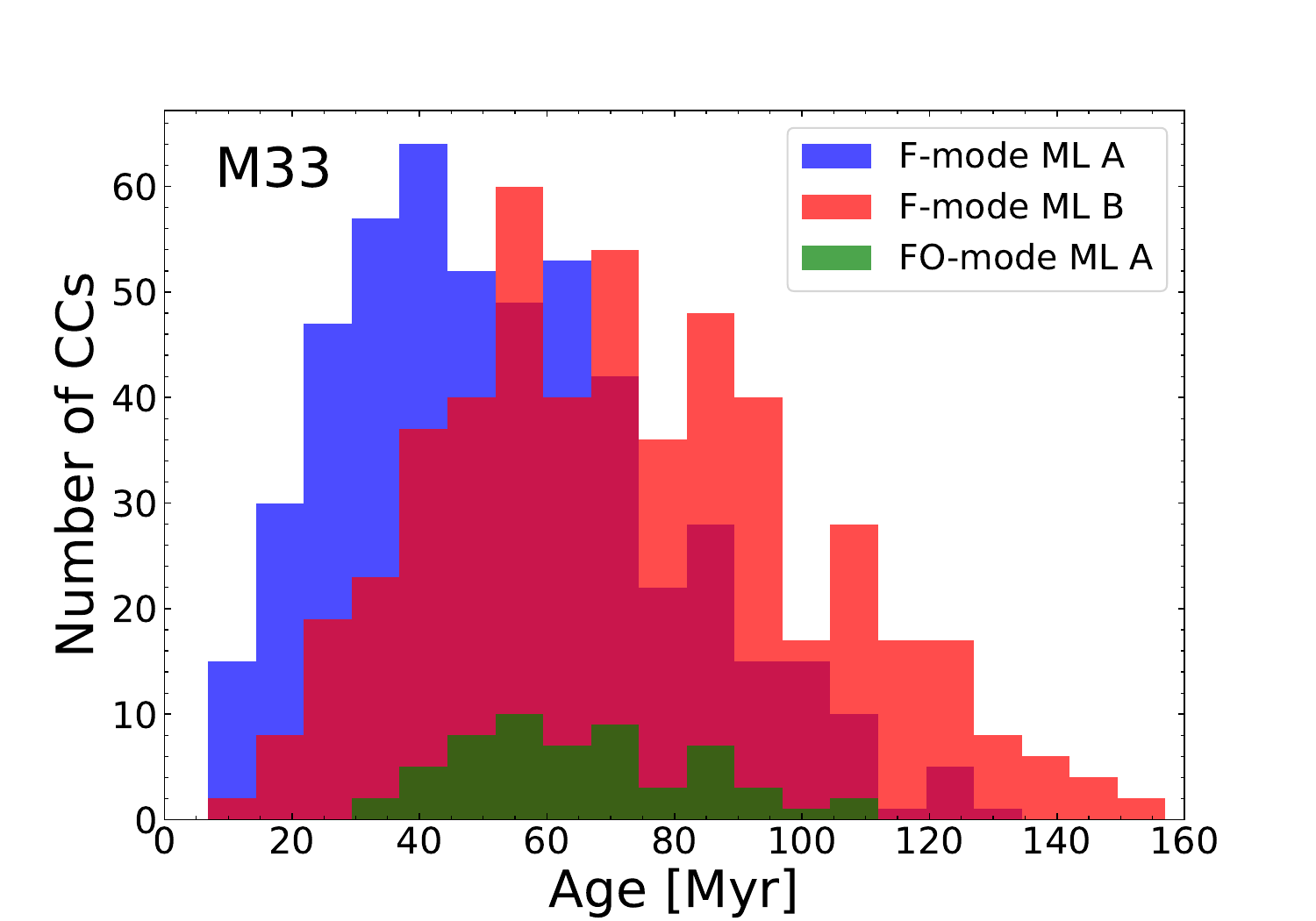}
    }
    \caption{\label{fig:hist} The predicted age distributions of CCs in the spiral galaxies M31 (left panel) and M33 (right panel), derived by applying the PAZ relation to selected samples of F-mode and FO-mode CCs. Left panel: blue, red, and green histograms show F-mode CC with ML A, F-mode CCs with ML B, FO mode CCs with ML A in M31, respectively. Right panel: same as left panel but for M33. Note that different scales are used in the left and right panels to account for the significantly larger CC sample in M31 with respect to M33.}
\end{figure*}

Figure \ref{fig:hist} presents the age distributions for M31 (left panel) and M33 (right panel), highlighting the dependence of these distributions on the pulsation mode (F-mode vs. FO-mode) and the assumed ML relation. Given the limited number of FO-mode models with ML B, \citetalias[][]{Desomma2021} provided only the PAZ relation for the canonical ML case. Inspection of the plots reveals that the non-canonical ML relation (case B), shifts the age distribution toward older ages. This finding is in agreement with previous studies, including \citet[][]{Anderson2016}, \citet[][]{Bono2005}, and \citetalias[][]{Desomma2020mnras}.

For the M31 CC age distribution (left panel of Figure \ref{fig:hist}), the canonical ML relation (case A) yields an age range for F-mode pulsators between 7.6 Myr and 154.9 Myr, with a peak at 68.7 Myr (blue bars). On the other hand, the non-canonical ML relation (case B) results in a broader age range, from 13.5 Myr to 170.2 Myr, with the peak shifting to 76.9 Myr (red bars). Canonical FO-mode pulsators, characterized by smaller masses and shorter periods compared to F-mode pulsators, exhibit an age distribution (green bars) clustered around older ages. Their ages range from 38.5 Myr to 119.3 Myr, peaking at 68.7 Myr.

For the M33 CC age distribution (right panel of Figure \ref{fig:hist}), the canonical ML relation produces an age range of 6.8 Myr to 131.4 Myr, peaking at 40.6 Myr. The non-canonical ML relation expands this range to 13.5 Myr to 157.1 Myr, with the peak shifting to 55.6 Myr. The FO-mode pulsators in M33, under the canonical ML assumption, show an age range of 32.8 Myr to 108.6 Myr, with a peak at 55.6 Myr.
In general, the F-mode ML A age distribution spans a younger range compared to the F-mode ML B and FO-mode ML A distributions. FO-mode pulsators consistently display narrower age ranges and older age peaks, reflecting their shorter periods and smaller masses.

In general, the F-mode age distribution for brighter ML relations (non-canonical models) yields systematically older ages. This is because non-canonical ML relations correspond to stellar models that account for core overshooting, which results in a larger convective core during hydrogen burning. Consequently, these models remain on the main sequence of the Hertzsprung-Russell diagram for a longer time compared to models that do not include overshooting. When these stars eventually evolve into CCs, they reach the instability strip at older ages.

Conversely, FO-mode pulsators consistently exhibit narrower age ranges and older age peaks, reflecting their shorter periods and lower initial masses.

We note that the inferred upper age limit for CCs in both galaxies (approximately 140 Myr in the canonical case and 170 Myr in the non-canonical case) may be affected by the lowest period limit of 2 days, as mentioned in Section 2.1.
To estimate the possible CC age range that may be missing in this analysis, we adopt the minimum period value of F-mode models used to derive the PAZ relation (P $\sim$ 1 day) and assume an average [Fe/H] value for F-mode Cepheids. Under these assumptions, we estimate ages of $\sim$180 Myr for the canonical case and $\sim$190 Myr for the non-canonical case.

This suggests that the present analysis does not provide information about CC ages in the range of 140 Myr to 180 Myr for the canonical case and 170 Myr to 190 Myr for the non-canonical case.

We note that the inferred age distributions do not represent actual star formation histories, as no initial mass function is assumed in the model computations, and different evolutionary times are taken into account only in the derivation of theoretical PAZ and PACZ relations.

The retrieved ages and their errors for M31 and M33 CCs are partially listed in Tables \ref{tab_age_31} and \ref{tab_age_33}

\begin{table*}
\centering
\caption{\label{tab_age_31} Individual ages for the F- and FO-mode M31 CCs in our sample, obtained using the canonical and non-canonical PAZ relations. The columns are: Galaxy name, source identifier, RA (in decimal degree), Dec (in decimal degree), pulsation mode, period (in days), age (in Myr) derived from the canonical (A) ML relation, error on the age (in Myr) derived from the canonical (A) ML relation, age (in Myr) derived from the non-canonical (B) ML relation, and error on the age (in Myr) derived from the non-canonical (B) ML relation. The full table is available in its entirety as a machine-readable table.}
\begin{tabular}{cccccccccc}
\hline\hline
Galaxy & Pan-STARRS1 Source Id & RA[deg] & DEC[deg] & Mode & P [days] & $t_{MLA}$[Myr] & $\sigma$ $t_{MLA}$[Myr] & $t_{MLB}$[Myr] & $\sigma$ $t_{MLB}$[Myr]  \\
\hline\hline
M31 & PSO J011.2795+41.1489 & 11.279 & 41.149 &    F &   1.971 &       154.979 &            29.624 &       170.729 &            31.951 \\
   M31 & PSO J010.5268+41.5728 & 10.527 & 41.573 &    F &   2.832 &       116.439 &            22.257 &       132.261 &            24.752 \\
...\\
 M31 & PSO J010.6222+41.6037 & 10.622 & 41.604 &   FO &   1.921 &       100.901 &             9.701 & - & - \\
   M31 & PSO J010.7287+41.0043 & 10.729 & 41.004 &   FO &   1.915 &       100.125 &             9.626 & - & - \\
...\\
\hline\hline
\end{tabular}
\end{table*}

\begin{table*}
\centering
\caption{\label{tab_age_33} Same as in Table \ref{tab_age_31} but for M33 galaxy.}
\begin{tabular}{cccccccccc}
\hline\hline
Galaxy & Source Id & RA[deg] & DEC[deg] & Mode & P [days] & $t_{MLA}$[Myr] & $\sigma$ $t_{MLA}$[Myr] & $t_{MLB}$[Myr] & $\sigma$ $t_{MLB}$[Myr]  \\
\hline\hline
M33 & J013243.21+302653.2 & 23.180 & 30.448 &    F &   2.483 &       131.470 &            25.130 &       157.009 &            29.384 \\
   M33 & J013454.32+305929.4 & 23.726 & 30.991 &    F &   2.640 &       126.011 &            24.086 &       154.431 &            28.901 \\
...\\
M33 & J013425.57+310029.2 & 23.607 & 31.008 &   FO &   2.052 &       108.649 &            10.446  & - & - \\
   M33 & J013319.01+302512.3 & 23.329 & 30.420 &   FO &   2.096 &       104.509 &            10.048  & - & -  \\
...\\
\hline\hline
\end{tabular}
\end{table*}

\section{Results}

\subsection{The density Maps of CCs in M31 and M33}

\begin{figure*}[ht]
\centering
    \vbox{
    \hbox{
    \includegraphics[width=0.45\textwidth]{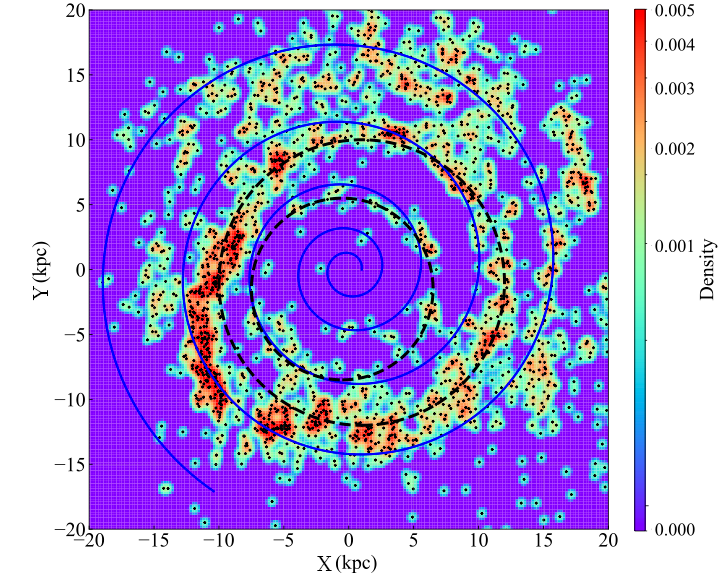}
    \includegraphics[width=0.45\textwidth]{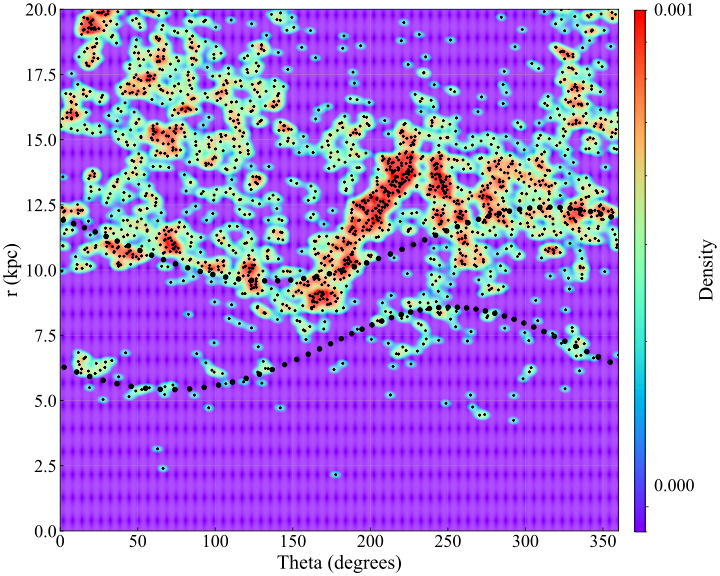}
    }
    \hbox{
    \includegraphics[width=0.47\textwidth]{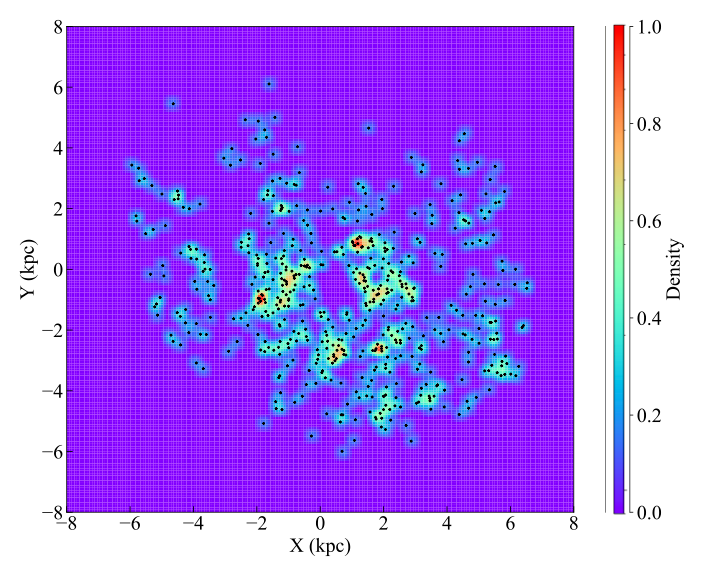}
    \includegraphics[width=0.46\textwidth]{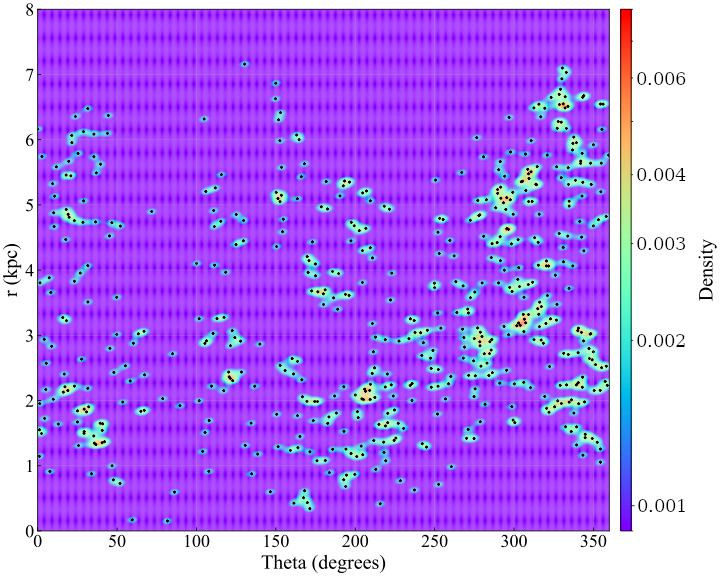}
    }}
    \caption{\label{fig:density_maps} Density maps of CCs in M31 (top panels) and M33 (bottom panels). Left panels: Spatial density distributions of CCs, smoothed using a Gaussian kernel density estimation.  In M31, a well-defined ring structure (dashed line) is visible at $\sim$ 11 kpc, consistent with previous studies, and a newly identified inner ring (dashed line) at $\sim$ 7 kpc is also observed.  Right panels: The same density maps shown in polar coordinates, highlighting ring-like structures in M31 (dotted curves) and their spatial distribution.}
    
\end{figure*}

First, we calculated deprojected Cartesian $X,\, Y$ coordinates (in kpc) for each CC in M31 and M33. To this aim, we used the following centers, distances, and viewing angles: RA=00:42:44.3 Dec=+41:16:7.5 D=773 kpc, PA=37.715 deg, $i$=77.5 deg for M31 \citep[data taken from][]{Zaritsky1994}; RA=01:33:33.1 Dec=+30:39:18
D=840 kpc, PA=22 deg, $i$=53 deg for M33 \citep[data taken from][]{Magrini2007}. We then calculated the $X,\, Y$ coordinates using the formulation by \citet{Wagner-Kaiser2015}. These rectangular coordinates were then used to calculate the polar radial and angular coordinates $R$ and $\theta$.  
As a first application, in the left panels of Figure \ref{fig:density_maps}, we present the density maps of CCs in M31 (upper panels) and M33 (lower panels). These maps are smoothed with a Gaussian kernel density estimation to reveal key structural features in the spatial distribution of CCs in both galaxies.

For M31, consistently with previous studies \citep[][]{Gordon2006, Kodric2013}, the map highlights the presence of well-defined CC concentrations along the spiral arms and identifies the presence of a ring pattern slightly off-centred ($X\, Y=1, -1$ kpc) and with radius $\sim$ 11 kpc, while previous studies \citep[e.g.][]{Kodric2018} estimated a radius of about 10 kpc.  
 
The possible explanations for the presence of this ring in M31 are thoroughly discussed in \citet{Block2006}, \citet{Gordon2006}, and \citet{Dier2014}. The most widely accepted explanation for its formation, proposed by these authors, is that it resulted from a past interaction with M32, one of M31’s satellite galaxies.

\citet{Block2006} and \citet{Gordon2006}, using infrared observations from the Spitzer IRAC (Infrared Array Camera) and MIPS (Multiband Imaging Photometer for Spitzer), respectively, concluded that the 10 kpc ring in M31 is a prominent star-forming structure and a dominant feature in M31’s disk. The ring appears nearly circular but exhibits a noticeable split near the location of M32. Moreover, M31’s spiral structure is weak and fragmented, suggesting an external perturbation rather than an internally driven process.

Both \citet{Block2006} and \citet{Gordon2006} performed dynamical simulations to test the hypothesis that a close passage of M32 through M31’s disk could have triggered the formation of the ring. Their models demonstrated that M32’s passage induced a density wave, compressing gas into a ring-like structure and leading to enhanced star formation at a radius of approximately 10 kpc. The simulations successfully reproduced the observed offset and splitting of the ring, further reinforcing the idea that M32’s gravitational influence played a crucial role.

Similarly, \citet{Dier2014} confirmed that a single off-center passage of M32 triggered a density wave in M31’s disk, inducing spiral arm-like perturbations, which, when projected onto the sky, appear as "pseudo-rings". Their simulation also showed that the interaction perturbed the velocity field of M31, leaving measurable kinematic effects still observable today. Notably, \citet{Dier2014} suggested that the M31–M32 interaction occurred approximately $\sim 800$ Myr ago.

However, earlier models by \citep{Block2006, Gordon2006} proposed a more recent collision, occurring approximately $\sim 200$ Myr ago. Since such a dynamical interaction is expected to trigger recent star formation, the CC age distribution derived in this work, which provides Cepheid ages not older than $\sim 200$ Myr, appears to support a more recent interaction between M31 and M32 than the timescale suggested by \citet{Dier2014}.

This finding underscores the relevance of studying CC age distributions as a tool to trace recent dynamical interactions among galaxies.

External to this structure, we observe a spiral arm (blue solid line) where the CC density increases, in agreement with previous literature findings. We have tentatively fitted this arm with a simple logarithmic function with a pitch angle of 320 degrees.

In addition to these already known structures, we note, for the first time, an internal ring having a center at $X\, Y=-0.5, -1.5$ kpc and radius $R\sim$ 7 kpc. 
The structure discussed above and particularly the inner ring are more clearly visible when plotted in polar coordinates, as shown in the top-right panels of Figure \ref{fig:density_maps}.

We speculate that the origin of this internal ring can be the same as that at $R\sim$ 11 kpc. Indeed, both are off-centered in the same direction (at least along the $Y$-axis) and show similar ages (see Figure \ref{fig:age_maps}).
To investigate this further, we defined two circular sectors centered on each ring, each with a radial width of 2 kpc.
We then calculated the average age of the Cepheids within these two sectors (104 and 307 Cepheids in the inner and outer rings, respectively, obtaining $<Age_{Inner}>=46\pm16$ Myr and $<Age_{Outer}>=53\pm19$ Myr. Since these ages are consistent within their uncertainties, our results support the idea that the two rings formed at similar epochs.
Therefore, our best current hypothesis is that the inner ring may have also formed as a result of M32's passage.

Contrary to M31, the density map distribution of the CCs in M33 exhibits a clumpier distribution of CCs without prominent spiral structures, suggesting a different star formation history and dynamical evolution. Also, the polar representation (see bottom right panel of Figure \ref{fig:density_maps}) does not provide any clue about possible substructures in the CC distribution.

\subsection{The CC Age distribution in M31 and M33}

\begin{figure*}[ht]
    \centering
    \vbox{
    \hbox{
    \includegraphics[width=1.0\linewidth]{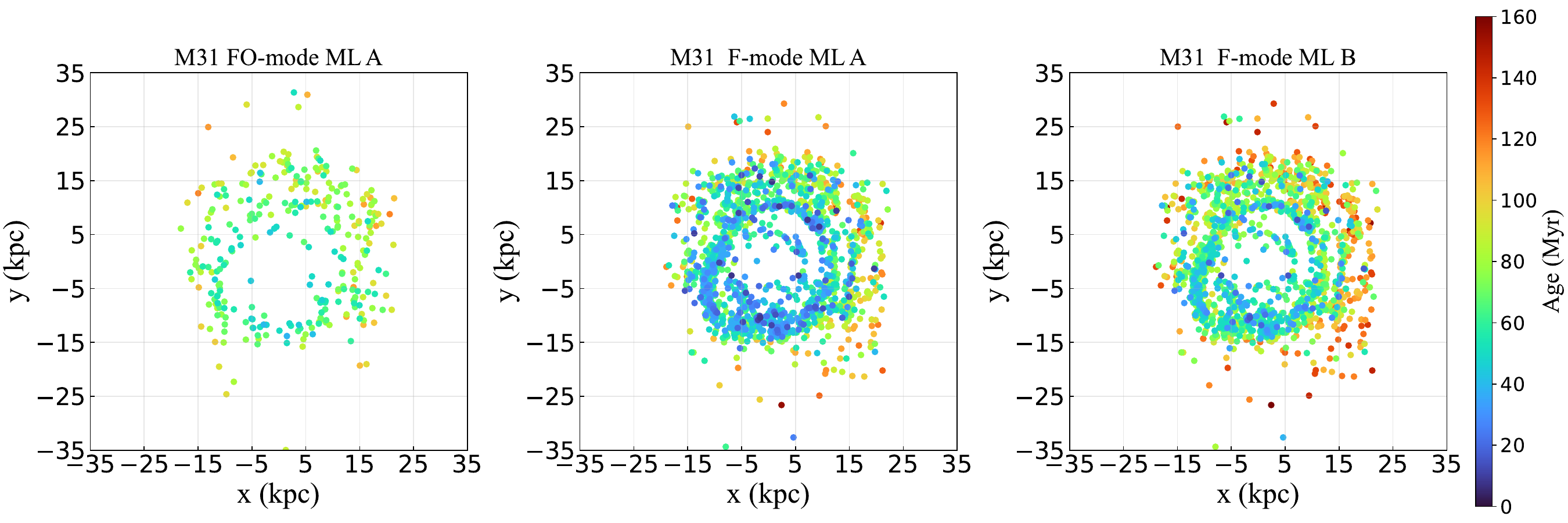}
    }
    \hbox{
    \includegraphics[width=1.0\textwidth]{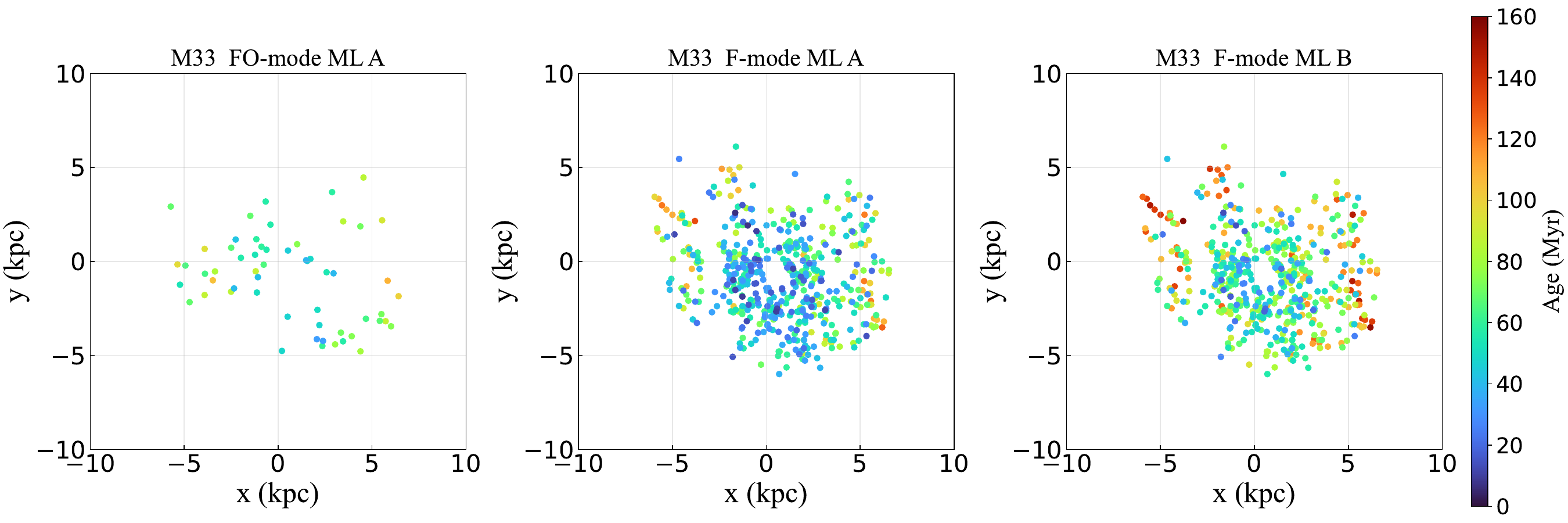}
    }}
    \caption{\label{fig:age_maps} Upper panels: Age maps of CCs in M31 from the PAZ relation, showing spatial age distribution for different pulsation modes and ML relations: (left) FO-mode (canonical A), (center) F-mode (canonical A), (right) F-mode (non-canonical B). The color bar indicates age in Myr. Bottom panels: Same as above for M33.}
\end{figure*}

By combining the calculated individual ages with the spatial distribution of F-mode and FO-mode CCs in M31 and M33, we infer the spatial age distributions of these galaxies. Using the individual ages listed in Tables \ref{tab_age_31} and \ref{tab_age_33}, we derived the age maps shown in Figure \ref{fig:age_maps}. These maps clearly illustrate the impact of the pulsation mode and ML assumptions on the derived age distributions of CC populations in M31 and M33.

A detailed inspection of these maps suggests that younger star formation episodes have been concentrated toward the inner disk, while older Cepheids are more abundant at larger radii. This trend seems to be a radial progression of star formation.
To further investigate age gradients, we examined their variations with the angular polar coordinates. For M31, we divided the map into four quadrants using a counterclockwise segmentation (see Figure \ref{fig:age_gradients_M31} in the appendix): 0° - 90°, 90° - 180°, 180° - 270° and 270° - 360°. For this analysis, we adopted ages derived from the canonical PAZ relation for the combined sample of F and FO pulsators.
The figure clearly shows the age trend between the galaxy's center and periphery. To be more quantitative, we derived linear regressions of the age distributions in each quadrant, fixing the intercept to a common value of 40.6 Myr (the intercept must be the same at the center of the galaxy). The resulting slopes in each quadrant are consistent within 1 $\sigma$. 
The residuals of the best-fitting lines are shown in Figure \ref{fig:plot_residui} in the Appendix. They exhibit different trends, some of which appear significant at more than the 1 $\sigma$ level. Particularly notable is the reversed trend observed in the 90°–180° quadrant compared to the others. We tentatively interpret this as the result of asymmetric star formation activity or dynamical effects.\\
In the case of M33, we found no evidence of variation in the age distribution with the angular polar coordinates, whereas we observed a significantly steeper global age gradient compared to M31 (refer to Figure \ref{fig:age_gradients_M33} in the appendix for details).

This suggests a higher concentration of young stars toward its center, reinforcing the idea that M33 has experienced a more centrally concentrated star formation activity compared to M31. This occurrence could be linked to the previously mentioned close interaction between M31 and M32, which had significant consequences, including the formation of the inner and outer rings. 
We emphasize that the age trends discussed above are not critically dependent on the adopted metallicity gradient. Indeed, we replicated the analysis using the simpler PA relation from \citet[][]{Desomma2020mnras}, i.e., neglecting the effect of metallicity on the Cepheid ages. As a result, the observed age trends are not significantly altered.

As mentioned in Sections 2.1 and 2.2, the possible lack of detection of CCs with periods shorter than 2 days may result in significant incompleteness in the oldest age bin. However, given the tightness of the age gradients shown in Figures \ref{fig:age_gradients_M31} and \ref{fig:age_gradients_M33}, it is unlikely that this missing bin would significantly alter the results.

\section{Discussion, Interpretation and Future Work}

In this study, we applied the updated PAZ relation from \citetalias[][]{Desomma2021} to derive individual ages for CCs in the spiral galaxies M31 and M33. 

By relying on the extensive CC catalogs from the PAndromeda survey \citep[][]{Kodric2013, Kodric2018} for M31 and the M33 Synoptic Stellar Survey \citep[][]{Pellerin2011} for M33, combined with metallicity gradients from \citet[][]{Zaritsky1994} and \citet[][]{Magrini2007}, respectively, we constructed spatial age distributions for these galaxies.

The resulting age distributions reveal clear age gradients across both galaxies, with an evident radial progression of star formation that may provide insights into galactic evolution. 
Furthermore, our findings confirm that the choice of the ML relation significantly influences the derived age distributions. The brighter, non-canonical ML relation, which accounts for a moderate amount of core convective overshooting (or mass loss or rotation) consistently shifts the age distributions toward older ages compared to the canonical relation.

For M31, we observed younger CCs concentrated towards the Galactic Center, while older populations dominate at larger galactocentric radii. Similar trends were observed in M33, albeit with a more pronounced age gradient, reflecting the galaxy's distinct metallicity gradient and star formation properties.

One of the most significant findings of this study is the confirmation of the well-known outer ring at $\sim$11 kpc in M31, which was previously identified in the literature and attributed to past interactions with M32 \citep[][]{Block2006, Dier2014, Gordon2006}. This structure aligns with previous interpretations of M31’s dynamical history, reinforcing the idea that tidal interactions have played a fundamental role in shaping its current morphology.

More remarkably, our analysis reveals an internal ring at $\sim$7 kpc for the first time. The origin of this newly identified structure remains an open question. One possible explanation is that it resulted from a localized episode of star formation, potentially driven by gas compression in the inner disk. We also notice that, given that both rings appear to have similar ages and are off-centered in the same direction, it is plausible that the internal ring also formed as a consequence of M32’s passage, suggesting that this interaction affected not only the outer regions of M31 but also its inner disk structure.

Our findings provide additional support for a more recent formation of the rings at $\sim$200 Myr ago, consistent with the earlier models proposed by \citet[][]{Block2006} and \citet[][]{Gordon2006}. Since such a dynamical interaction is expected to trigger recent star formation, the Cepheid age distribution derived in this work, with no CCs older than $\sim$200 Myr, aligns with the scenario of a more recent interaction between M31 and M32 than the timescale suggested by \citet[][]{Dier2014} ($\sim$800 Myr ago).

This study represents the first application of the PAZ relation to Local Group spiral galaxies and highlights the utility of CCs as stellar population tracers, demonstrating how age distributions can be used to reveal recent dynamical interactions in galaxies.

Further investigation is needed to determine whether the internal ring is dynamically induced or a result of internal star formation, requiring spectroscopic analysis of young stellar populations and gas kinematics. Future studies will extend this analysis to other galaxies, incorporating more precise metallicity measurements and refining pulsation and evolutionary models underlying the PAZ derivation.

In particular, future observations with the Nancy Grace Roman Space Telescope \citep[Roman, see e.g.][]{Faddaroman25} and the James Webb Space Telescope \citep[JWST, see e.g.][]{Gardner2006, Rigby2023} will provide significant advancements in studying CC populations in external galaxies. Roman’s wide-field imaging will enable precise measurements of CC distances and ages across large galactic regions, reducing crowding effects and improving sample completeness. Meanwhile, JWST’s infrared sensitivity will be particularly valuable for detecting CCs in dusty star-forming regions, where optical observations are limited by extinction.

Both telescopes will contribute to a better understanding of star formation histories and help refine the impact of metallicity on CC evolution, further improving the calibration of period-age-metallicity relations.

\clearpage

\appendix 

\section{The Age Gradients of M31 and M33}

\begin{figure*}[ht]
\centering
\includegraphics[width=\linewidth]{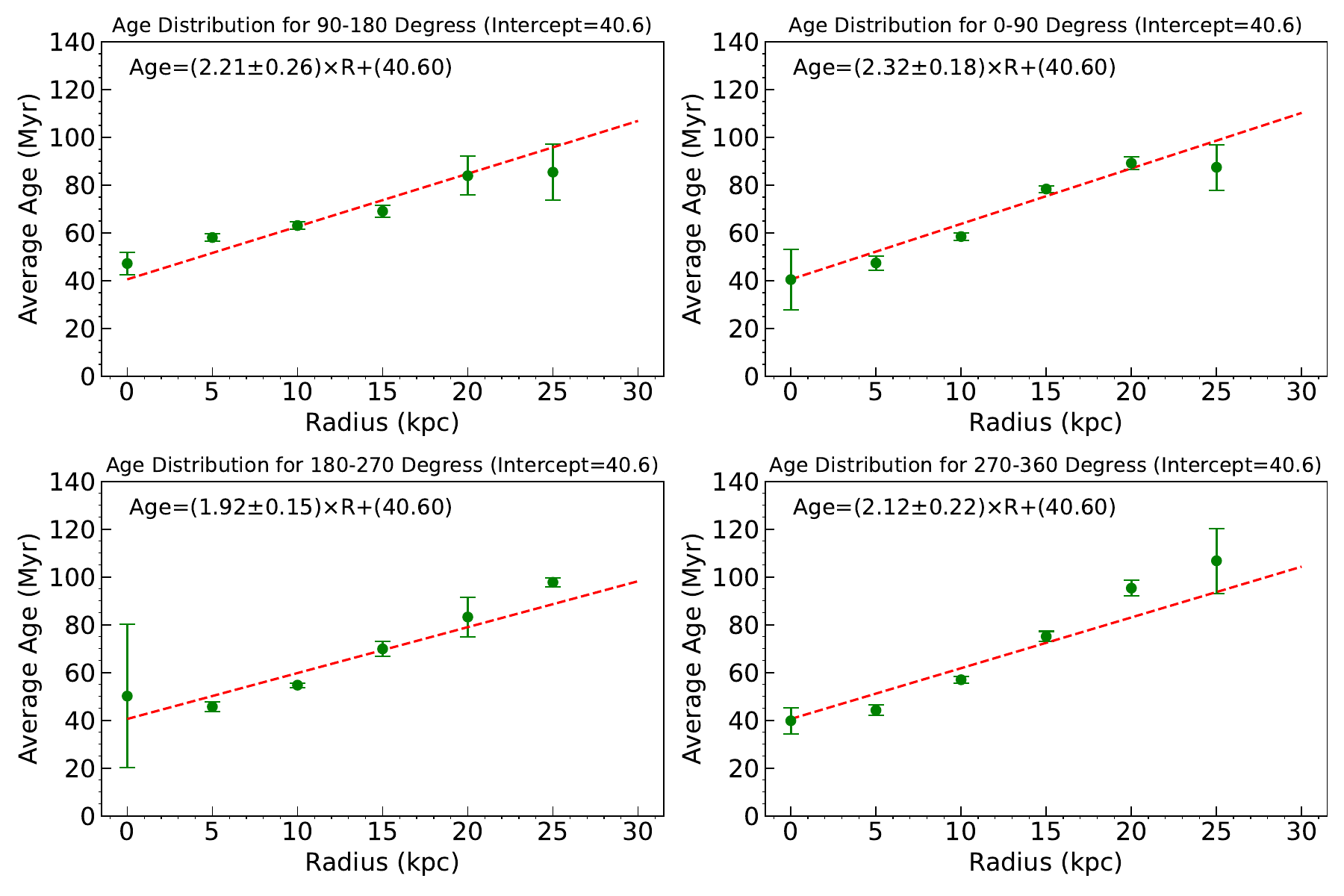}
\caption{\label{fig:age_gradients_M31} Age gradients of CCs in M31 across four quadrants of the galaxy: 0°–90° (top right), 90°–180° (top left), 180°–270° (bottom left), and 270°–360° (bottom right). Each panel shows the average age of CCs as a function of galactocentric radius (in kpc), with linear fits overlaid performed with a fixed intercept at 40.6 Myr.} The blue dots represent the mean ages of CCs in radial bins, while the red dashed lines indicate the best-fit linear regression models. For reference, the disk scale length of M31 has been estimated to be $\sim$ 5.5 kpc \citep[][]{Yin2009}.
\end{figure*}

\begin{figure*}[ht]
\centering
\includegraphics[width=\linewidth]{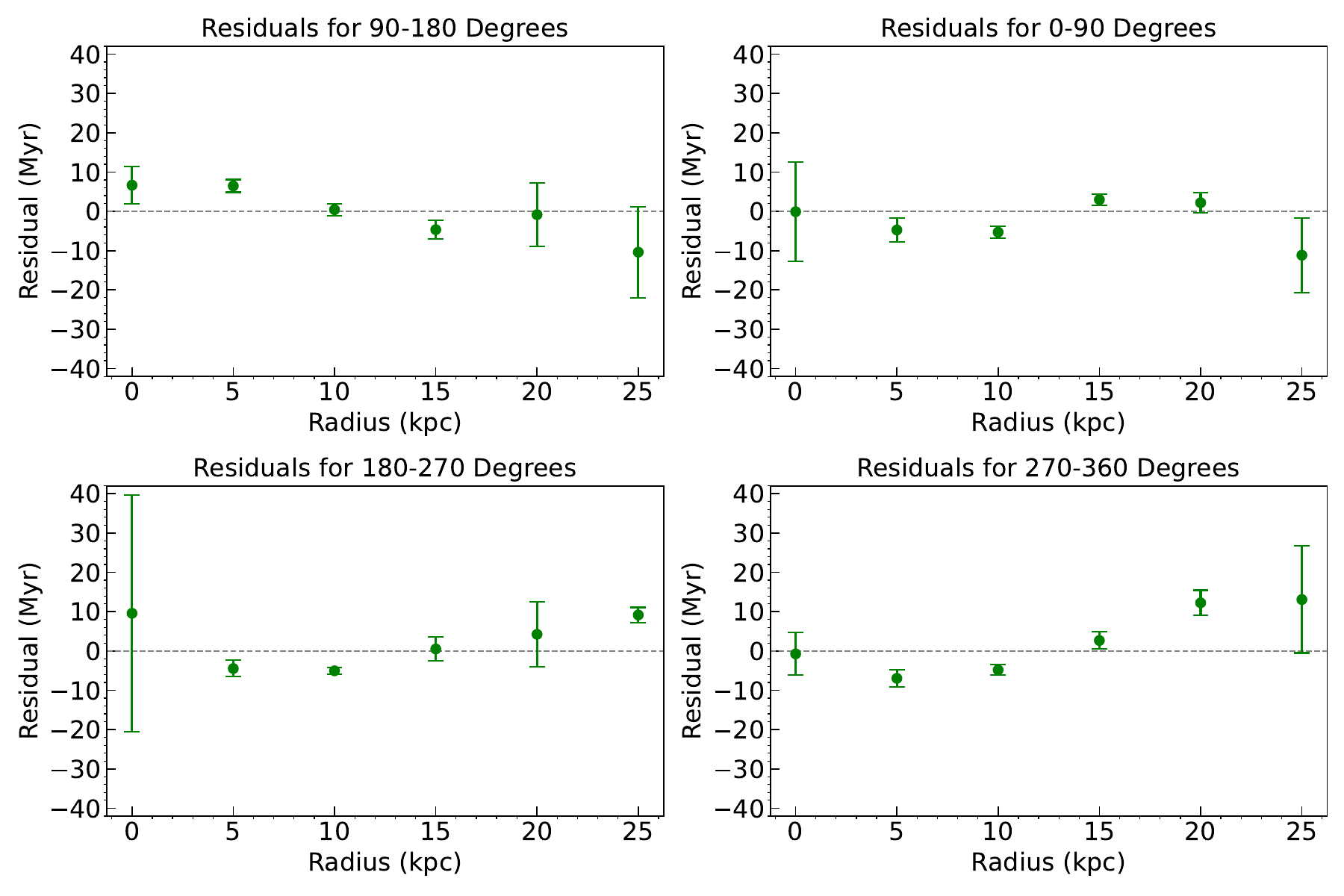}
\caption{\label{fig:plot_residui} Residuals between observed and model-predicted average ages for CCs in each galactic quadrant, based on a linear fit with a fixed intercept of 40.6 Myr. }
\end{figure*}

\begin{figure*}[ht]
\centering
\includegraphics[width=0.5\linewidth]{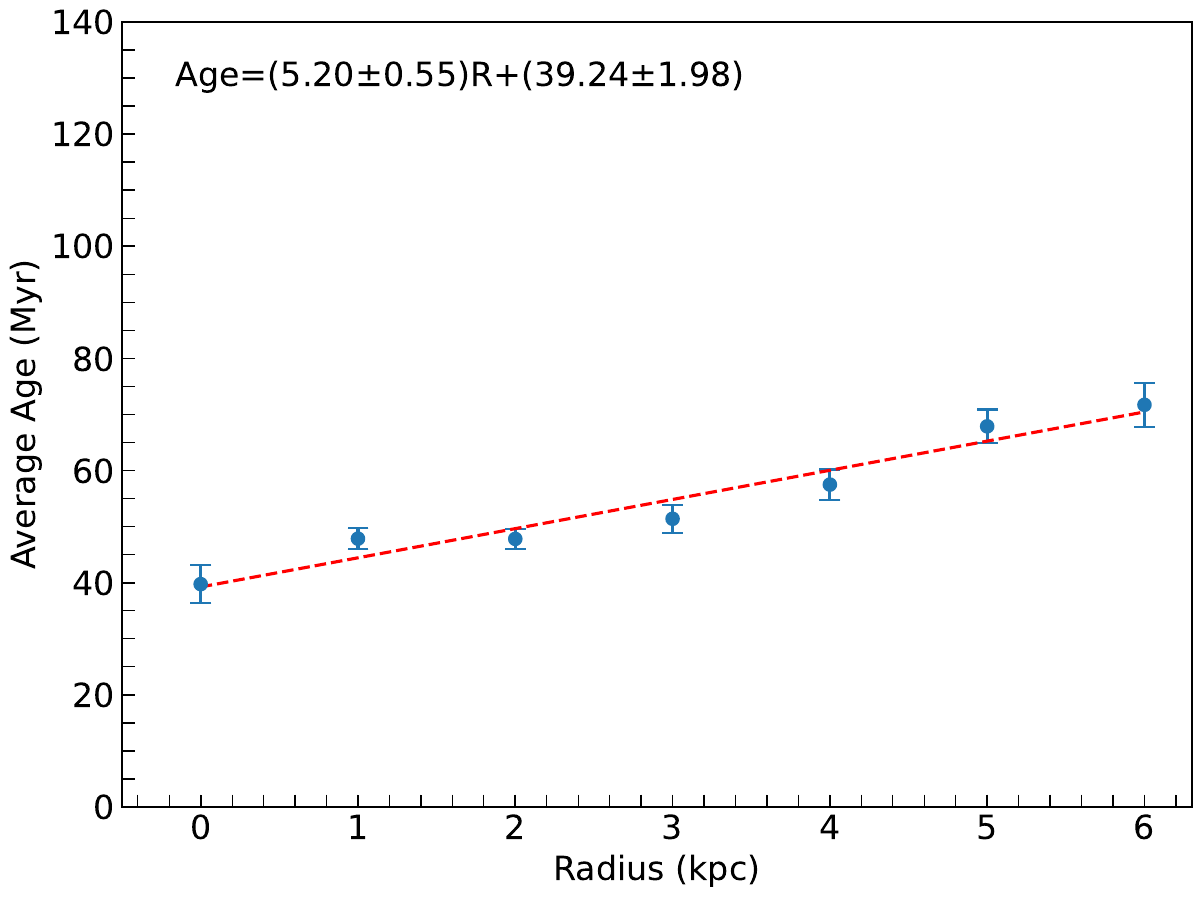}
\caption{\label{fig:age_gradients_M33} Global age gradient of CCs in M33. as a function of galactocentric radius (in kpc). Blue dots indicate the mean ages of CCs in radial bins, while the red dashed line represents the best-fit linear regression model. 
For reference, the disk scale length of M33 has been estimated to be $\sim$ 4.34 kpc \citep[][]{Smercina2023}.}
\end{figure*}

\section*{Acknowledgements}
We appreciate the Referee’s insightful comments, which have greatly contributed to improving the clarity and robustness of our manuscript.
This project has received funding from the INAF-ASTROFIT fellowship, the PRIN MUR 2022 project (code 2022ARWP9C) "Early Formation and Evolution of Bulge and Halo (EFEBHO)," PI: M. Marconi, funded by the European Union – Next Generation EU, and the Large Grant INAF 2023 MOVIE, PI: M. Marconi.

We acknowledge funding from Gaia DPAC through INAF and ASI (PI: M.G. Lattanzi) and the INAF GO-GTO grant 2023 "C-MetaLL – Cepheid Metallicities in the Leavitt Law" (PI: V. Ripepi).

G.D.S. and T.S. thank INFN (Naples section) for support via QGSKY initiatives, with additional INFN support for Moonlight2 (G.D.S.).

M.M. and V.R. acknowledge ISSI for funding the project "EXPANDING" (ISSI–ISSI Beijing Team).

SC acknowledges the support of a fellowship from La Caixa Foundation (ID 100010434) with fellowship code LCF/BQ/PI23/11970031 (P.I.: A. Escorza) and from the Fundación Occident and the Instituto de Astrofísica de Canarias under the  Visiting Researcher Programme 2022-2025 agreed between both institutions.

This research also benefited from COST Action CA21136 (CosmoVerse), addressing cosmological tensions through systematics and fundamental physics, funded by COST (European Cooperation in Science and Technology)."**

\bibliography{desomma_main_apj}{}
\bibliographystyle{aasjournal}

\end{document}